\documentclass[%
reprint,
superscriptaddress,
 amsmath,amssymb,
 prl,
]{revtex4-2}

\usepackage[american]{babel}
\usepackage{graphicx}% Include figure files
\usepackage{dcolumn}% Align table columns on decimal point
\usepackage{bm}% bold math
\usepackage{color}
\newcommand{\tm}[1]{\textcolor{black}{#1}}
\newcommand{\jad}[1]{\textcolor{black}{#1}}

\begin{document}

%\preprint{APS/123-QED}

\title{Confinement controls the creep rate in soft granular packings}
%Creep without imposed fluctuations in a soft particle packing
%Jamming me softly with his density: creep flow across a rigidity transition in hydrogel sphere packings.
%Intruder creep in athermal soft particle packings close to jamming
%lyrics line: Everybody got a breakin' point

\author{Joshua A. Dijksman}
 \email{j.a.dijksman@uva.nl}
 \affiliation{Van der Waals-Zeeman Institute, Institute of Physics, Science Park 904, 1094KS, Amsterdam, The Netherlands}
 
\author{Tom Mullin}
 \email{tom.mullin@maths.ox.ac.uk}
\affiliation{The Mathematical Institute and Linacre College,
 University of Oxford.OX2 6GG,U.K.
}

%Potential reviewers
%Susan Fielding
%Olivier Pouliquen
%Doug Durian
%Eric Clement
%Daniel Bonn

\date{\today}% It is always \today, today,

\begin{abstract}
Flow in soft materials encompasses a wide range of viscous, plastic and elastic phenomena which provide challenges to modelling at the  microscopic level. To create a controlled flow, we perform falling ball viscometry tests on packings of soft, frictionless hydrogel spheres. Systematic creep flow is found when a controlled driving stress  is applied \jad{to a sinking sphere embedded in a packing}.  Here, we  \tm{ we take the novel approach of applying an additional global confinement stress to  the packing using an external load. This has enabled} us to \tm{identify two distinct}  creep regimes. When confinement \jad{stress is} small, the creep rate is independent of the  load imposed. For \jad{larger} confinement stresses, we find that the creep rate is set by the mechanical load acting on the packing. \jad{In the latter regime, the} creep rate depends exponentially on the imposed stress. We \jad{can} combine the two  regimes \jad{via a rescaling onto a master curve}, capturing the creep rate over \jad{five} orders of magnitude. Our results indicate that bulk creep phenomena in \tm{these soft} materials can be \tm{subtly} controlled using \tm{an external mechanical force}.
\end{abstract}

\maketitle

%\tableofcontents

\maketitle

\section*{Introduction}
The \tm{physical} properties of athermal particle packings have a variety of  non-trivial features which are of interest at both the fundamental and applied levels. Collections of materials  such as sand, foams, emulsions  and other particulate media have a ``rigid'' phase that can bear a finite amount of stress~\cite{evans_concentration_1990, bartsch_effect_2002, hecke_jamming_2009, shao_role_2013, vlassopoulos_tunable_2014, coussot2014yield, basu2014rheology, villone_dynamics_2019, o2019jammed, shewan2021viscoelasticity}. However, the definition of ``rigid'' is sometimes not clear cut since slow mechanical motion or \emph{creep} can exist in thermally driven amorphous materials~\cite{andrade1910viscous,spaepen1977microscopic}. Specifically, packings of inelastic particles might be considered rigid, yet they also display slow relaxation dynamics when \emph{deformation} is imposed, even in the absence of thermal fluctuations; they are considered to self-fluidize~\cite{PhysRevLett.81.2934,PhysRevLett.103.036001}. Alternatively, when \emph{stress} is imposed, granular packings also display very small magnitude logarithmic aging~\cite{darnige2011creep, deshpande2021perpetual}. Hence, the origin of creep in athermal packings is unclear. Despite the introduction of concepts such as non-locality, there is no general framework to connect the microscopic details of (a)thermal particle packings and their fluctuations to their rheology at a coarse grained level. Furthermore, it is difficult to reproduce these observations numerically  using \tm{techniques such as} Discrete Element Methods without introducing  (ad hoc) noise.

In this study we focus on elastic phenomena using packings of hydrated hydrogels. Previous work~\cite{2022creepcontrol} on this material has revealed  systematic dependence of creep flow rates on driving stress. Here we show that the creep in athermal soft particle packings can also be controlled via an external \emph{mechanical} stress applied to the particle phase.  We  again find that the local stress that the intruder applies to the packing exponentially enhances the creep rate. We can merge these two competing effects using  a master curve, in which a stress-time superposition principle is captured. Our work highlights sensitivity to boundary stresses of bulk creep phenomena.

\begin{figure}[t]
\centering
  %\includegraphics[width = 8cm]{interface2.pdf}
  %\caption{}
  %\label{fig:interface}
  \includegraphics[width = 9cm]{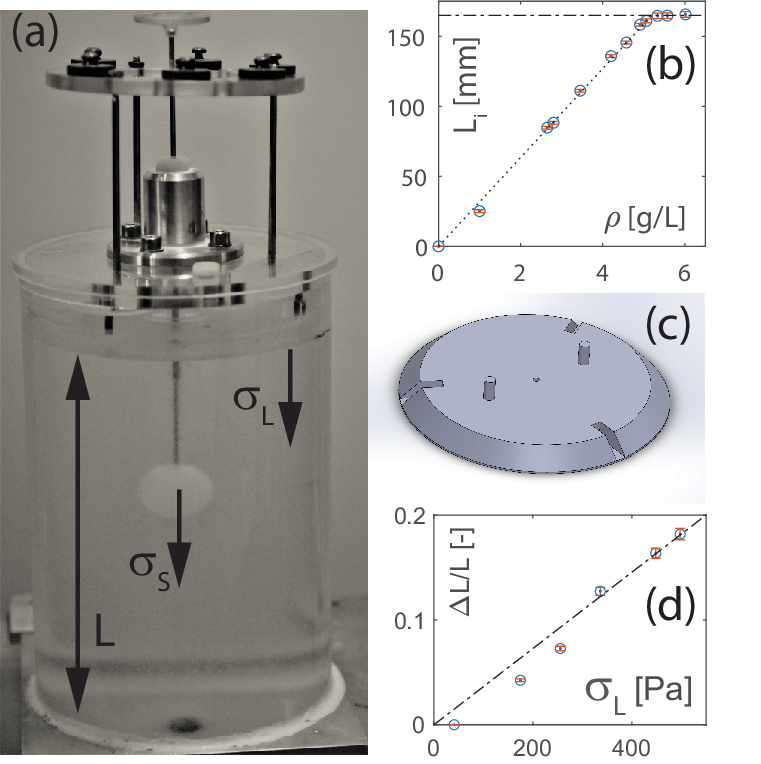}
  \caption{(a) Photo of the setup with the guided lid, used for larger confinement stresses. $L$ indicate the lid position from the base. $\sigma$ is the stress applied to the packing. \jad{(b) The initial height of the hydrogel layer $L_i$ versus the concentration $\rho$ of dry hydrogel powder added to $2$ Litres of water. (c) Drawing of the 3D printed lid used to provide low confinement stress to the hydrogel packing. The protrusions are for picking up the lid.} (d) Confinement stress $\sigma_L$ compresses the packing by amount $\Delta L$ as indicated by the data. The dashed straight line with a slope of $3.3$\,kPa to guide  the eye.}
  \label{fig:setup}
\end{figure}
\section*{Experimental setup}
\begin{figure}[t]
\centering
\end{figure}

%\paragraph{This is the next level heading.~~} For this level please use \texttt{\textbackslash paragraph}. These headings should also end in a full point.
\begin{figure}[t]
\centering
  \includegraphics[width = 8cm]{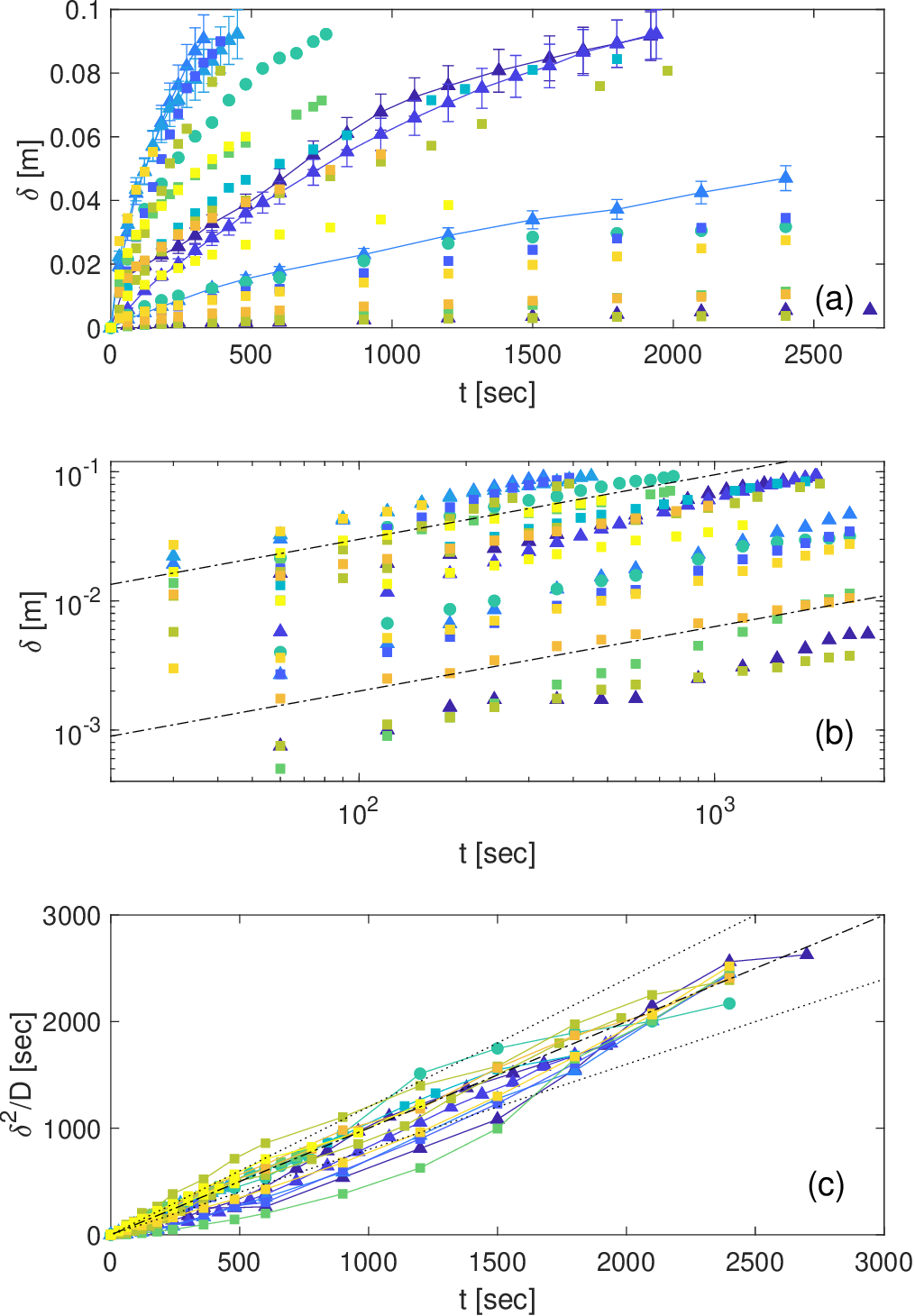}
  \caption{(a) Selection of sinking data for experiments without lid ($\circ$), unsupported lid ($\triangle$) and the supported lid ($\square$). \jad{The colors identify the evolution of $\delta$ in a single experiment, and indicate different experiments at different, arbitrarily selected settings.} Error bars are indicated for line-connected, unsupported lid data but representative for all measurements. The same data is shown on log scale in (b), with trend lines $\delta \propto t^{1/2}$ indicated as dashed line.  (c) The squared displacement normalized by fit coefficient $D$ as a function of time. The dash-dotted line has a slope of 1. The dotted lines have a slope of 20\% smaller and larger, to indicate the error on $D$.}
  \label{fig:pheno}
\end{figure}

We prepare a packing of hydrogel spheres using a successfully established approach~\cite{2022creepcontrol}. Briefly, we swell hydrogel beads (JRM Chemicals, type ``snow'') in \tm{ a fixed volume of $2$ Litres of }triply boiled Oxford tap water. The particles are swollen in a $124$mm diameter Plexiglass container - see Fig.~\ref{fig:setup}a. \tm{ The dry particles are sprinkled into the water in measured amounts of $\approx 2$ gms. They are heavier than the water and sink and swell to form a layer of swollen particles at the bottom of the container. A gentle stir of the layer is applied between additions of particles and the layer is left for $2$ hours to settle between additions. Interestingly, it is found that the depth of the layer is linearly dependent on the weight of the dry beads as shown in Fig.  \ref{fig:setup}(b). This observation provides evidence for the uniformity of the density of the layer.}  \tm{ We find it convenient to define the volume fraction $\rho = \frac{m}{V}$ where $m$ is the weight of the dry hydrogel powder and $V$ is the volume of water.} \tm{The fixed volume of  $2$ Litres of water used in our experiments, corresponds to a water level at $L= 165$mm. \tm{At the specified} concentration of hydrogel powder (and no additional confinement of the packing), the packing reaches the surface of the water i.e. there is no clear water above the hydrogels - see Fig.~\ref{fig:setup}b.} 
We apply confinement stress by placing a rigid lid on the surface of the packing. The lid is always submersed so that \tm{there is no influence of surface tension  on  the particle packing.} For the first set of experiments, a lid was designed and 3D printed in plastic (PLA), partially hollow and weighing $\simeq 50.3$gram as shown in Fig.~\ref{fig:setup}c. The lid rests freely on the surface of the packing; when submersed, its contribution to the stress on the particles is negligible.  We added weights to the upper surface of this free lid (FL) to increase the confinement stress $\sigma_L$. The lid remained level when small masses were added but it tended to tilt when adding more than $\sim300$ grams in calibrated weights to the lid. We therefore designed a second, guided lid (GL) for subsequent experiments as shown in Fig.~\ref{fig:setup}a. In this setup, the same container \jad{is} used, but the lid is now guided by three rods which pass through closely fitting holes in the cover of the container. The three rods are connected by a $2$ mm thick ring on which we can add weights to increase the confinement stress - see Fig.~\ref{fig:setup}d. \jad{We also observe that the amount of compression of the packing $\Delta L/L$ increases approximately linearly with the amount of applied \jad{confinement} stress, indicating a packing stiffness of about 3.3 kPa as indicated in Fig.~\ref{fig:setup}d. The measured packing stiffness is lower than that of the particles themselves (about 10kPa), which is reasonable as the packing is a loose, porous collection of soft, deformable and slippery particles.} The rods  ensure that the lid remains horizontal over the range of confinement stresses investigated.  All experiments above $200$ Pa confinement stress level are achieved with the GL, which allows us to  apply a  stress of up to $\simeq 1000$ Pa.\\
\jad{We probe the mechanical behavior of the prepared packing via what is essentially a falling ball viscometry test. Inside the packing of swollen hydrogel spheres, we embed an intruder. The intruder is a spherical plastic object, 2\,cm in diameter and mounted on a rod. The rod is guided by a vertical bearing to maintain \tm{the alignment} of the rod and allow for tracking of the intruder motion, while the intruder can sink into the packing of spheres \tm{as a result of} the gravitationally induced stresses it exerts on the packing}. The intruder stress $\sigma_{S}$ \jad{that the intruder is exerting} is varied by adding calibrated weights to the tray which is connected to the $23$~mm diameter plastic sphere by a thin rod. The 3D printed tray alone  weighs $2$ grams which enables stresses down to $\simeq$ $100$ Pa to be applied. We compute the intruder stress from the buoyancy corrected total weight of the intruder, the volume of the sphere and the surface of the sphere. We also test sinking rates without the lid (NL) to verify reproducibility of previous results~\cite{2022creepcontrol}. We collect displacement data via photographs taken at set intervals and post-processing of the images to extract the penetration depth $\delta(t)$ as a function of time. \jad{We measure $\delta$ from the starting position of the sphere, which is always fully submersed in the packing.}\\ 
We  found it to be essential to stir the packing, and allow a settling period of three hours between rehearsals of the experiment as in previous work. Stirring the packing is carried out carefully to avoid introducing bubbles, as the presence of bubbles \tm{creates} poorly reproducible sinking dynamics. We stir the packing by burying a long rod at the bottom of the container. The rod is nearly as wide as the container and has two neodymium magnets at its ends. The magnets are held with other magnets on the outside of the container. \tm{ Slow movement of the external magnets allows us to change the position of  the rod within the packing and thereby induce stirring while keeping the lid in place}.

\section*{Experimental Results}
The main experimental results for the  sinking characteristics of the sphere are presented in Fig.~\ref{fig:pheno}a. \jad{\tm{Note that only one third of  all the data collected over about one year of experimental efforts are shown here. The examples are randomly selected and displayed in this short form for reasons of clarity whereas the full dataset is used in the analysis}. The color coding spans different experimental settings in $\sigma_L$ and $\sigma_S$ and \tm{its purpose is to} highlight the time dependence $\delta(t)$}. For all three cases NL, FL, GL, and a wide variety of $\sigma_{S}, \sigma_L$, we observe the characteristic non-linear displacement dynamics as \tm{reported in our} previous experiments. The data is well represented by square root behavior, as observed previously and indicated in Fig.~\ref{fig:pheno}\jad{a} and on log scale in Fig.~\ref{fig:pheno}\jad{b}. The error on the determination of the position is small, as indicated. Consistently, we  find that $\delta^2(t) \propto t$ as evidenced by the collapse of data in Fig.~\ref{fig:pheno}c. Here we divide out the fitted prefactor $D$  from $\delta^2 = Dt$ to find that all data follows a single master curve. We  observe small deviations from linearity in the $\delta^2$ vs $t$ plot, but see no systematic trend in the deviations with any experimental control parameter.

\subsection*{Intruder stress dependence}
\begin{figure}[t]
\centering
  \includegraphics[width = 8cm]{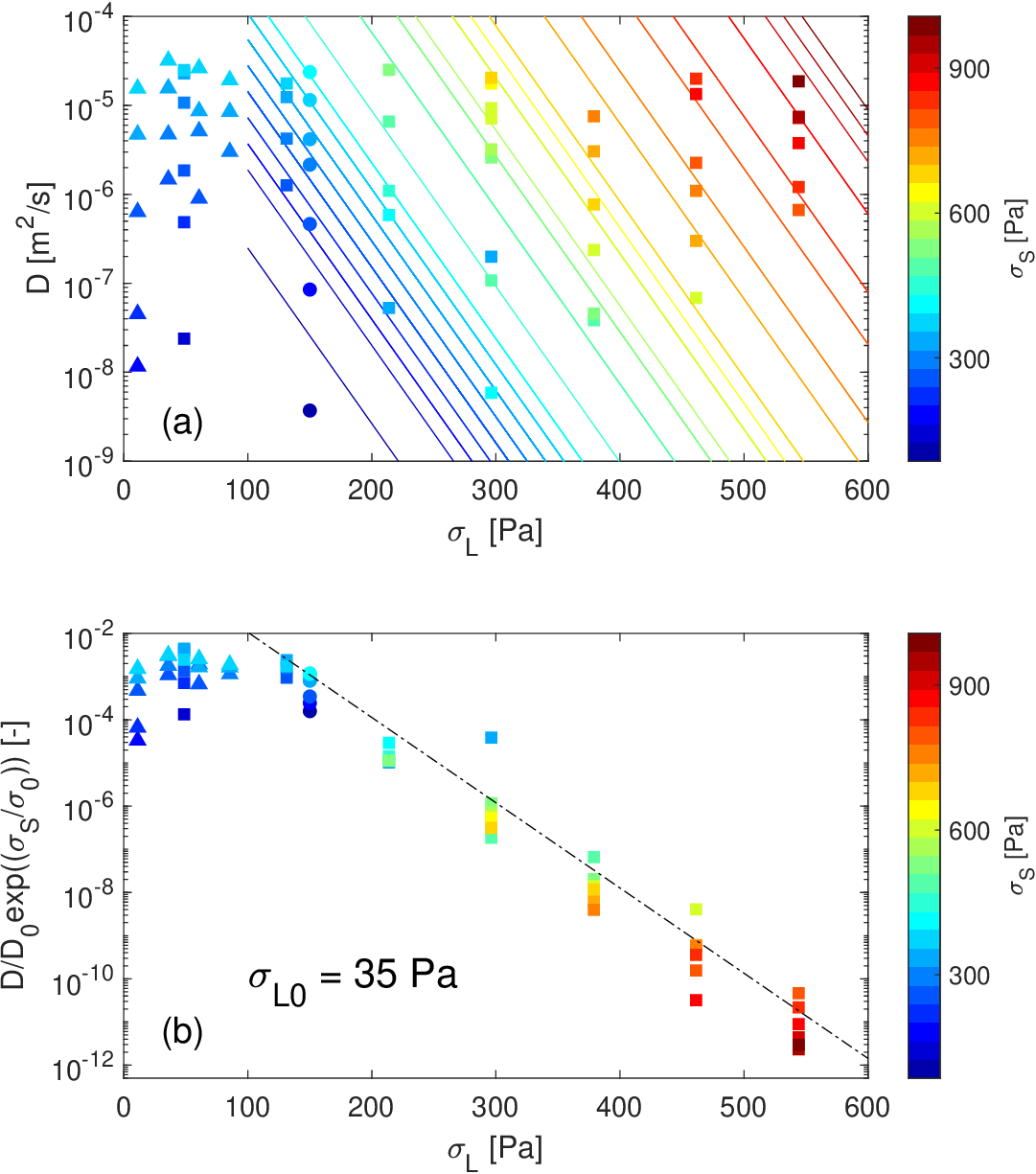}
  \caption{(a) All $D$ obtained for all $\sigma_{L,S}$ measured shown versus confinement stress. Symbols indicate experiments NL ($\circ$), FL ($\triangle$) and GL ($\square$). Solid lines indicate the trend as described in Eq.~\ref{eq:sigmaSLdep} (b) Rescaled data. The dashed line indicates the exponential decay with confinement stress; the stress scale $\sigma_{L0}$ is indicated in the panel.\label{fig:surfstress}}
  
\end{figure}

We measure the sinking rate $D$ for a range of intruder stresses $\sigma_{S}$ and confinement stresses $\sigma_L$. Having access to all $D(\sigma_{S},\sigma_L)$, we can establish the stress dependence via two different representations of the data. We first discuss how $D(\sigma_{L})$ depends on $\sigma_S$. 
We show all measured $D(\sigma_{S},\sigma_L)$ in Fig.~\ref{fig:surfstress}a. Given the quality of the collapse of Fig.~\ref{fig:pheno}c, we conclude that the error bars on the slope estimate $D$ is less than 20\% and hence becomes negligible on the log scale representation used henceforth. There are three main observations: \textit{(i)} at low confinement stress $\sigma_L$, $D$ \jad{appears to be} approximately independent of the confinement stress: both FL and GL data here \tm{suggests there is a}  a plateau in $D(\sigma_L)$; the \jad{tentative} plateau value  depend on $\sigma_{S}$. \textit{(ii)} The data of the NL case are consistent with the FL and GL case. This consistency indicates that current experiments probe the same dynamics as previous 
work~\cite{2022creepcontrol} in which no lid was used, \textit{(iii)} above approximately 100\,Pa of confinement stress, the creep rate has a definite dependence on the confinement stress. We find that the previously observed exponential stress dependence (shown in solid lines) captures the overall trend  well:
\begin{equation}
    D = D_0e^{\left(\frac{\sigma_{S}}{\sigma_{S0}} - \frac{\sigma_L}{\sigma_{L0}}\right)} \label{eq:sigmaSLdep}
\end{equation}
Here, $\sigma_{S0,L0}$ are characteristic stress scales for the intruder and confinement stress respectively. The dependence of $D\propto\exp{(-\sigma_L/\sigma_{L0})}$ is \tm{immediately} obvious from the slope of the solid lines following the trends in all data in Fig.~\ref{fig:surfstress}a. The dependence of $D\propto\exp(\sigma_{S}/\sigma_{S0})$ is indicated by the vertical separation of the colored solid lines; also  the matching of \tm{the} functional dependence of $D(\sigma_{S0},\sigma_L)$ and the data is evident. 
Further evidence that the creep rate of the intruder depends exponentially on the stresses involved, is provided by rescaling $D$ with its purported exponential intruder stress dependence. A result of this rescaling is the collapse of the data shown in Fig.~\ref{fig:surfstress}b. Let us call $\Tilde{D} = D/\exp(\sigma_{S}/\sigma_{S0})$. We see again that \textit{(i)} the high stress data collapses onto a single master curve $\Tilde{D} \propto \exp(-\sigma_L)$. 
\tm{Furthermore, data taken in} the low stress regime \tm{provides a reasonable} collapse onto a single line, \jad{suggesting} that the low stress regime is  governed by a single $\sigma_{S0}$, despite its independence of confinement stress. We find that for $D_0 = 2\times 10^{-6}$, $\sigma_{S0} = 35$\,Pa, $\sigma_{L0} = 22$\,Pa, all data collapses onto a single confinement stress dependent curve, suggesting that the bulk creep rate is indeed set by the mechanical boundary stress acting on the particles and competing with the local stress  from the intruder. It should be emphasized that we show here all the  data, from NL, FL and GL experiments, demonstrating consistency across  different lid types and hence ways to exert boundary stress. Note that to estimate the amplitude of the surface stress in the NL case, we need to make assumptions \tm{concerning} the typical radius of curvature of the water surface between the hydrogel beads. An estimate of about 100\,Pa was previously considered reasonable, but this value \tm{has} considerable uncertainty. As we can see, a value of 150\,Pa \tm{produces a good match between}  the NL data and the trend line; however a surface tension stress of 100\,Pa would also keep the NL data consistent with the FL and GL cases. \jad{We emphasize here that the visible change in the behavior of the hydrogels at $D(\sigma_{L})$ around 100\, Pa is therefore unrelated to the surface tension pressure scale, as we shall see with more careful analysis in the following section.}\\

\section*{Confinement stress dependence}
\begin{figure}[!t]
\centering
  \includegraphics[width = 8.5cm]{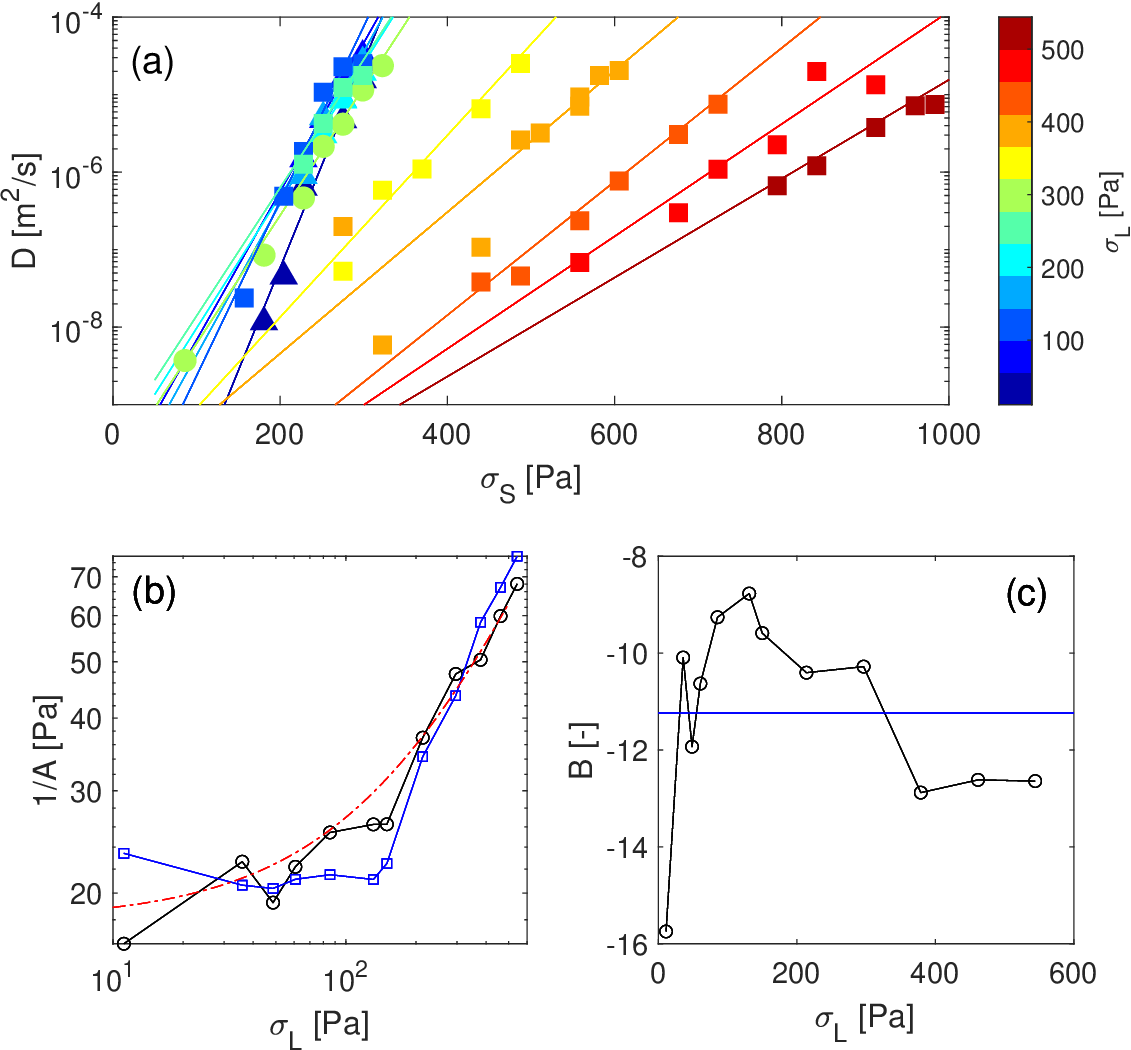}
  \caption{(a) All $D$ obtained for all $\sigma_{L,S}$ measured shown versus sphere stress. Symbols indicate experiments NL ($\circ$), FL ($\triangle$) and GL ($\square$). Solid lines indicate the trend of Eq.~\ref{eq:sigmaSLdep}. (b) Intruder stress scale factor $A$ fitted with free $B$ (black circles) or fixed $B=-11.2$ (blue squares). The red dash-dotted line represents $1/A \propto\sigma_L$ with $A_0=18$\,Pa and $k=0.09$ (c) $B(\sigma_L)$ from the fits (black circles) or fixed at their average fitted value $B=-11.2$ (blue line)}
  \label{fig:partstress}
\end{figure}

Even though the collapse displayed in Fig.~\ref{fig:surfstress}b is satisfying, small \jad{but systematic} deviations can be observed over the range of $\sigma_S$. Further,  the low confinement stress regime appears to be different\jad{ --- careful inspection of the data reveals that the  plateau  shows significant scatter over about two orders of magnitude, even in the rescaled $D$. In fact, scatter of about one order of magnitude is present in the rescaled $D$ values at higher values of $\sigma_{L}$}. We can \jad{gain a better understanding of the physical mechanisms at play} by plotting $D$ as a function of the intruder stress. We show $D(\sigma_{S}$) and its dependence on $\sigma_L$ in Fig.~\ref{fig:partstress}a. We observe that the FL, NL and \tm{the} GL observations with low $\sigma_L$ have approximately the same\jad{, but not identical} trend with $\sigma_{S}$. \jad{This trend with $\sigma_{S}$ is the reason why the rescaling with Eq.~\ref{eq:sigmaSLdep} as shown in Fig.~\ref{fig:surfstress} does not collapse onto a single master curve.} At larger confinement \jad{stress}, the intruder stress \jad{effect} shows dependence on confinement. Moreover, the magnitude of the confinement stress \jad{dependence} becomes larger with $\sigma_L$. To capture \tm{the} \jad{effects} of the change of slope \jad{in both low and high confinement stress regimes}, we fit all constant confinement stress data with a\jad{n additional \tm{compensatory}} linear function on a semilog scale. The fits are shown as solid lines. Specifically, we have fitted

\begin{equation}
    \log(D) = A(\sigma_L)\sigma_S + B(\sigma_L),\label{eq:sigmaLdep}\\
\end{equation}

It is instructive to consider the behavior of the fitted prefactors $A,B$. We show their dependence on $\sigma_L$ in Fig.~\ref{fig:partstress}b,c. The prefactor $1/A$ sets the slope of the intruder stress dependence \jad{on the creep rate} and hence is essentially a pressure scale. It is therefore natural to assume that $1/A \propto \sigma_L$, and the data confirm this. The results of two \jad{different fitting} methods to extract $1/A$ are shown in Fig.~\ref{fig:partstress}b,c. One can keep $B$ free to have a $B(\sigma_L)$, or fix it at some $B$. Both methods of extracting $A$ yield the dependence to be well-fitted by $1/A = A_0 + k\sigma_L$, with $A_0=18$\,Pa and $k=0.09$ (red dashed line).

\section*{Discussion}
The different portrayals of $D$ in Eqs.~\ref{eq:sigmaSLdep} and \ref{eq:sigmaLdep} can be seen as two different representations of the same behavior. Indeed, we can understand that the weak $\sigma_L$ dependence  makes Eq.~\ref{eq:sigmaLdep} \jad{the more broadly applicable description than} Eq.~\ref{eq:sigmaSLdep}, as we found that

\begin{eqnarray}
    D = e^{\left(A(\sigma_L)\sigma_S+B\right)},\\
    = e^Be^{\left(\frac{\sigma_S}{A_0 + k\sigma_L}\right)},\\
    \simeq e^Be^{\left(\frac{\sigma_S}{A_0} - \frac{k\sigma_S\sigma_L}{A_0^2}\right)},
\end{eqnarray}

\begin{figure}[!t]
\centering
  \includegraphics[width = 9cm]{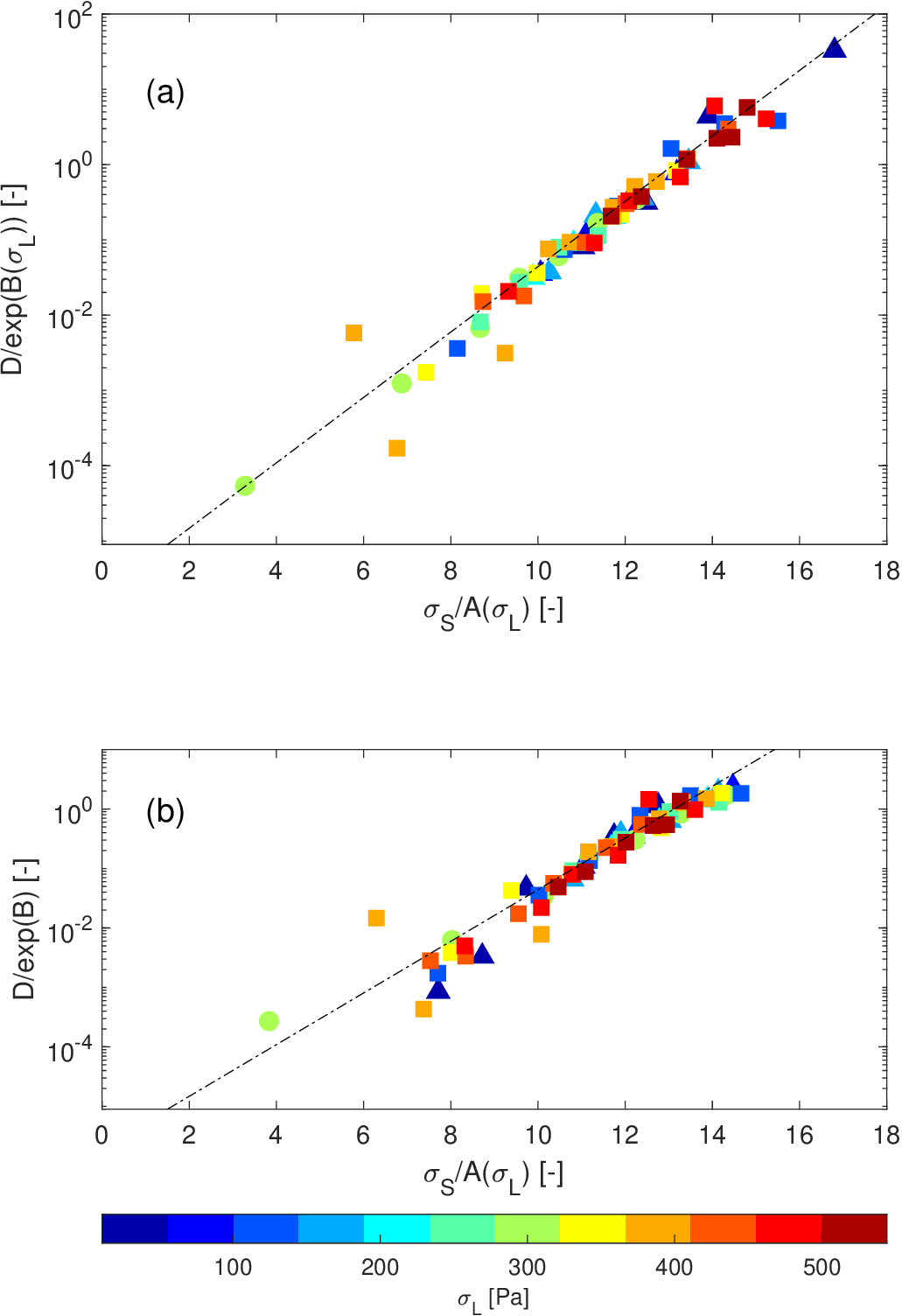}
  \caption{(a) Rescaled data, dividing out the $\sigma_L$ dependence \jad{for the freely fitted Eq.~\ref{eq:sigmaLdep}, providing $\sigma_{L}$-dependent $A$ and $B$. (b) Rescaled data, now with the fixed $B=-11.2$ while fitting $A(\sigma_{L})$ as shown as blue squares in Fig.~\ref{fig:partstress}b. The colorbar applies to both panels.} The dashed line indicates the exponential intruder stress dependence $D_0\exp(\sigma_S)$ in both panels.}
  \label{fig:fincoll}
\end{figure}

In the limit of $\sigma_L < \sigma_S$ explored in \cite{2022creepcontrol}, the weak confinement stress dependence appears to be the balance of stresses used to construct Eq.~\ref{eq:sigmaSLdep}. There, $\sigma_{S0} \approx \sigma_{L0}$ and also in Eq.~\ref{eq:sigmaSLdep} we see that $A_0 \approx k\sigma_S/A_0^2$ for $\sigma_S \approx 100$\,Pa. To further highlight the \jad{compatibility} of the two equations, we can scale out the dependence of both the stress variables. The creep rate $D$ can be corrected for the confinement stress by plotting $D/\exp(B(\sigma_L))$, while the driving stress $\sigma_S$ can be normalized by $A(\sigma_L)$. The collapse of the data over more than six orders of magnitude is visible in Fig.~\ref{fig:fincoll}a: clearly there is a universal rate-stress superposition principle at play here, in which simply the rescaling factors depend on the applies stresses. Also, in the limit of $\sigma_S \rightarrow 0$, the intercept for $D \approx D_0 \approx e^B$ as expected. The offset $A_0$ of about 18\,Pa \jad{is the \tm{most likely} source of the qualitatively different creep behavior observed at low confinement stresses, and} has a physical interpretation: besides the confinement from the lid, the hydrogel particles experience a weak hydrostatic pressure gradient because \tm{the hydrogels} are not perfectly density matched. Our previous work estimated this pressure scale to be about 10\,Pa for the geometry and materials used. We thus conjecture that the finite slope \jad{in $\log(D(\sigma_{S})$} to which the creep data converges at low $\sigma_L$ \jad{as visible in Fig.~\ref{fig:partstress}a} is due to hydrostatic pressure effects.\\ 
The meaning of dimensionless factor $B$ is less obvious. \jad{The variation in $B$ is significant, \tm{particularly} as it appears in an exponential factor. Even so, any change in $B$, e.g. the apparant low value of $B$ at low $\sigma_L$, is somewhat offset by a change in the value of $1/A$, as visible when comparing Figs.~\ref{fig:partstress}b,c. As the effect of a changing $B$ can also be absorbed in an effective diffusion constant $D_0$, there are good reasons to assume that $B$ depends on the experimental settings. For example,} fluctuations in temperature or perhaps \tm{small mis-alignments of the lid} could potentially produce suck deviations in the prefactor. \jad{Pragmatically,} keeping $B$ fixed to the average value of freely fitted $B(\sigma_L)$ of course yields a lower, yet still satisfactory, quality fit and collapse of the data. \jad{We show this collapse in Fig.~\ref{fig:fincoll}b. We observe that} fixing $B$ does not essentially affect the quality of the rescaling. \jad{We can thus conclude that Eq.~\ref{eq:sigmaLdep}  captures \tm{very well} the creep response of the hydrogel packing.}

\section*{Conclusions}
We report a consistent set of results on creep behavior found using  a spherical intruder in a packing of hydrogel particles under a variety of confinement conditions. By varying the confinement stress on the packing, we observe that the boundary induced stress influences the creep rate. The creep is also observed to be dependent on the intruder stress, in agreement with published results. The creep rate depends exponentially on both local and global stress scales. This rate-stress dependence can be rescaled and collapses onto a master curve. The rescaling factors reveal a residual role for a small hydrostatic pressure gradient. Our results highlight the relevance of boundary stresses on creep dynamics.\\

\section*{Conflicts of interest}
There are no conflicts to declare.

\section*{Acknowledgements}
We acknowledge stimulating discussions at the Salamina workshop in Bari, and the feedback from Jorge Peixinho, Edan Lerner and Tommaso Pettinari. We appreciate the help of Raoul Fix in designing and constructing part of the setup. We are grateful to Keith Long who built the majority of the apparatus in Oxford. This project was completed in part by funding from the European Union’s Horizon 2020 research and innovation program under the Marie Skłodowska Curie grant agreement No 812638.

%%%REFERENCES%%%
%\bibliography{rsc} %You need to replace "rsc" on this line with the name of your .bib file
%\bibliographystyle{rsc} %the RSC's .bst file

\providecommand*{\mcitethebibliography}{\thebibliography}
\csname @ifundefined\endcsname{endmcitethebibliography}
{\let\endmcitethebibliography\endthebibliography}{}

\end{document}